# Reversibly Strain Engineering and Electric-Field Control of Crystal Symmetry in Multiferroic Oxides

Fei Sun[1], Chao Chen[1], Deyang Chen[1,a)], Minghui Qin[1], Xubing Lu[1], Xingsen Gao[1], Christopher T Nelson[2], Jun-Ming Liu[1,3]

[1]Institute for Advanced Materials and Guangdong Provincial Key Laboratory of Optical Information Materials and Technology, South China Normal University
Guangzhou 510006, China

[2]Materials Science and Technology Division, Oak Ridge National Laboratory,
Oak Ridge, Tennessee 37831, USA

[3]Laboratory of Solid State Microstructures and Innovation Center of Advanced Microstructures, Nanjing University, Nanjing 210093, China

[a)]Authors to whom correspondence should be addressed:
deyangchen@m.scnu.edu.cn

ABSTRACT

Multiferroic oxides, such as $BiFeO_3$ (BFO), have garnered significant attention due to their coupled ferroelectric, magnetic, and elastic properties, offering exciting opportunities for multifunctional device applications. Controlling phase transitions in these materials is critical for tuning their physical properties and achieving desired functionalities. While numerous studies have focused on ferroelectric-ferroelectric transitions at rhombohedral-tetragonal (R-T) morphotropic phase boundaries (MPBs), far less attention has been given to the ferroelectric-antiferroelectric phase boundaries. Such systems hold promise for discovering novel physical phenomena, such as reversible phase transitions, enhanced piezoelectricity, and magnetoelectric coupling. In this work, we report a reversible antiferroelectric-to-ferroelectric phase transition in La-doped $BiFeO_3$ (LBFO) thin films. By modulating the residual strain via film thickness, an antiferroelectric orthorhombic (O) phase is stabilized within a ferroelectric rhombohedral (R) phase matrix. Under an external electric field, the phase transitions reversibly between these two states. This discovery not only enriches the understanding of O/R morphotropic phase boundaries but also provides a potential pathway for developing magnetoelectric devices with enhanced functionality.

# INTRODUCTION

Epitaxial strain provides an effective and alternative pathway to induce phase transitions in complex oxide thin films.[1-4] For instance, compressive strain can drive the rhombohedral-to-tetragonal (R-T) phase transition,[5-7] while tensile strain induces the rhombohedral-to-orthorhombic (R-O) phase transition in $BiFeO_3$ (BFO) thin films.[8-10] In recent years, numerous intriguing physical phenomena have been observed near the R/T morphotropic phase boundary (MPB), such as enhanced piezoelectricity,[11] magnetism,[12] and photoconductivity.[13] Additionally, the R-to-T phase transition can be reversibly driven by electric fields, accompanied by large strain and a shape-memory effect.[14, 15]

Meanwhile, a variety of exotic physical properties near the O/R morphotropic phase boundary (MPB) have garnered significant attention, including enhanced piezoelectric and dielectric responses,[16, 17] magnetic properties,[18] conductivity,[19] ferroelectric photovoltaic effect,[20] and the O/R ferroelectric MPB associated with polar vortices.[21] These pioneering studies open new avenues for exploring O/R mixed-phase BFO thin films. However, much of the research on MPB systems has focused primarily on ferroelectric-ferroelectric phase boundaries, with little attention given to ferroelectric-antiferroelectric phase boundaries. Recently, our previous work[22] reported a strain-driven antiferroelectric-to-ferroelectric (O-R) phase transition in La-doped BFO thin films. It was observed that the orientation of the antiferromagnetic axis can shift from out-of-plane (R phase) to in-plane (O phase), enabling the modification of antiferromagnetic axes by driving the antiferroelectric-to-ferroelectric phase transition. Building on this, we fabricated O/R mixed-phase LBFO thin films to investigate the reversible electric-field switching of the antiferroelectric-ferroelectric (O-R) phase transition, which is expected to achieve magnetoelectric coupling.

In this study, we show that antiferroelectric orthorhombic (O) and ferroelectric rhombohedral (R) phase structures in LBFO thin films can be reversibly controlled through epitaxial strain and external electric fields. We find that epitaxial strain stabilizes the R-phase, while intermediate strains induce a coexistence of both the ferroelectric R phase and antiferroelectric O phase. The topography of O/R mixed-phase LBFO films displays a distinctive striped contrast, enabling direct observation of antiferroelectric-to-ferroelectric (O-R) phase transitions via surface displacements. Additionally, we observe a reversible electric-field-induced transition from the antiferroelectric (O) to ferroelectric (R) phase. Our results highlight that both epitaxial strain and external electric fields are effective tools for controlling phase transitions in LBFO thin films, making them promising candidates for magnetoelectric memory and antiferromagnetic spintronic applications.

## MATERIALS AND METHODS

We use a nominal composition of $La_{0.2}Bi_{0.8}FeO_3$ (LBFO) as the target material, which exhibits mixed rhombohedral (R) and orthorhombic (O) phase structures. LBFO thin films of varying thicknesses (19 nm, 45 nm, and 140 nm) were grown on (110)-oriented $DyScO_3$ (DSO) substrates (orthorhombic with pseudo-cubic lattice parameters of a = b = 3.944 Å, c = 4.005 Å; the subscript "O" refers to the orthorhombic structure) using pulsed laser deposition (PLD) with a KrF excimer laser ($\lambda$ = 248 nm). The growth temperature and oxygen pressure during film deposition were set at 690 °C and 100 mTorr, respectively. To study the antiferroelectric-to-ferroelectric (O-R) phase transition behavior of LBFO thin films under an electric field, a 20-nm-thick $SrRuO_3$ (SRO) (a = 3.94 Å) buffer layer was epitaxially grown as the bottom electrode. X-ray diffraction (XRD) and reciprocal space mapping (RSM) were employed to characterize the structural evolution with film thickness. Surface displacements and polarization switching behaviors were measured using atomic force microscopy (AFM) and piezoresponse force microscopy (PFM) as a function of the applied electric field. Additionally, transmission electron microscopy (TEM) was used for detailed structural characterization of the antiferroelectric and ferroelectric phases.

## RESULTS AND DISCUSSION

To investigate the direct effect of epitaxial constraint on the antiferroelectric-to-ferroelectric (O-R) phase evolution, we grew a series of epitaxial LBFO thin films on DSO substrates. Fig. 1a-c show the atomic force microscopy (AFM) topographic images of LBFO thin films with thicknesses of 19 nm, 45 nm, and 140 nm, respectively. The atomically smooth surface morphologies indicate high-quality epitaxial growth of the LBFO films. In Fig. 1a, the topography reveals a uniform height contrast with a terrace structure, suggesting that the 19 nm-thick LBFO thin film is in the R phase, stabilized by compressive epitaxial strain and exhibiting high-quality epitaxial growth. In contrast, the 45 nm-thick LBFO film (Fig. 1b) shows dark contrast stripes, which we attribute to the O/R mixed phase due to partial relaxation of the epitaxial strain. As strain is further released in the 140 nm-thick LBFO thin film (Fig. 1c), a significant amount of dark stripe-like O phase is observed embedded in the terraced R phase matrix. This behavior resembles the characteristic features of the R/T mixed phase commonly reported in BFO thin films.[5]

To better understand the phase structures, typical θ-2θ X-ray diffraction (XRD) patterns of LBFO thin films with thicknesses of 19 nm, 45 nm, and 140 nm are shown in Fig. 1d, confirming that all LBFO thin films are epitaxially grown on DSO substrates. The out-of-plane lattice parameter of the 19 nm-thick LBFO thin film, calculated from the position of the $002_{pc}$ peak (where "pc" refers to the pseudo-cubic structure), is c ≈ 3.961 Å, which closely matches the lattice parameter of the bulk BFO phase.[23] As the film thickness increases, an additional $002_{pc}$ diffraction peak appears at approximately 46.5°, indicating the emergence of the O phase in the 45 nm-thick LBFO thin film, with a pseudo-cubic lattice parameter of c ≈ 3.920 Å, consistent with earlier studies.[24] The peak shifts to a higher angle as the film thickness increases, demonstrating further strain relaxation in the 140 nm-thick LBFO thin film, which has a lattice constant of c ≈ 3.917 Å. Therefore, the AFM measurements and XRD patterns in Fig. 1 illustrate that epitaxial strain plays a crucial role in driving the O-R phase transitions.

To further investigate the evolution of phase structures, we performed detailed structural characterization and analyzed the strain state of LBFO thin films with varying thicknesses using reciprocal space mapping (RSM). The $(0\text{-}13)_{pc}$ diffraction peaks of the LBFO thin films, corresponding to the $(240)_O$ diffraction peaks of the DSO substrates, are shown in Fig. 2(a)-(c). These data confirm that all LBFO thin films are coherently strained to the DSO substrates.

Fig. 2(a) shows a single diffraction peak, indicating that only the R phase is present in the 19 nm-thick LBFO film. From the $(103)_{pc}$ diffraction peaks of the LBFO films shown in Supporting Information Fig. S1(a), we calculated the in-plane lattice parameters to be a ≈ b ≈ 3.946 Å, with relatively small compressive strain (-0.35%). In contrast, the RSM of the 45 nm-thick film (Fig. 2(b)) displays two distinct diffraction peaks, signifying the coexistence of R and O phases. Notably, the splitting peak of the O phase shifts along the $[-110]_O$ direction, indicating that strain relaxation drives the formation of the O phase in the LBFO thin film.[22] The lattice parameters of the O phase, extracted from Fig.. 2(b) and Supplementary Fig. S1(b), are a ≈ 3.955 Å and b ≈ 3.946 Å.

In Fig. 2(c), a further shift of the O phase diffraction peak is observed, signaling the complete strain relaxation in the 140 nm-thick LBFO thin film. The in-plane lattice parameters for this film were calculated to be a ≈ 3.958 Å and b ≈ 3.946 Å (Fig. S1(c)). A summary of all lattice parameters for the R- and O-phase LBFO thin films with different thicknesses is provided in Fig. S1(d), illustrating the evolution of the ferroelectric-antiferroelectric (R-O) phase structure as the epitaxial strain relaxes with increasing film thickness.

These RSM data are consistent with our XRD results (Fig. 1d), confirming that gentle epitaxial strain effectively controls R-O phase transitions in LBFO thin films. We propose that the ability to modulate phase stability using epitaxial strain is closely related to the composition of the LBFO target, which, due to its proximity to the R-O phase boundary, is particularly sensitive to strain.

To further investigate the nature of the crystal structures, we examined the 1/4(101) diffraction peaks of LBFO thin films with different thicknesses using reciprocal space mapping (RSM), as shown in Fig. 2(d)-(f). The emergence of quarter-order diffraction peaks is indicative of a structure consistent with an antiferroelectric phase.[24,25]

Focusing first on the 19 nm-thick LBFO thin film in Fig. 2(d), we observe only a (103)pc diffraction peak, which confirms that the film consists solely of the ferroelectric R phase. In contrast, in the 45 nm-thick LBFO thin film (Fig. 2(e)), a 1/4(101) superstructure spot (marked with arrows) appears in addition to the fundamental

(103)pc diffraction peak, signifying the presence of an antiferroelectric phase structure. As the film thickness increases further, another 1/4(101) superstructure diffraction spot becomes clearly visible in Fig. 2(f), confirming the emergence of the antiferroelectric phase in the 140 nm-thick LBFO thin film.

The uneven intensity distribution between the 1/4(101) superstructure diffraction spots is attributed to the preferred orientation of the film, which is influenced by the orthogonality of the DSO substrate.[26] These results, combined with the extracted lattice parameters (Fig. 2a-c), provide strong evidence that the R-O structural transition is a ferroelectric-to-antiferroelectric phase transition.

We grew a 100 nm-thick LBFO thin film on a DSO substrate with a 20 nm-thick SRO bottom electrode, in order to study domain structures, The X-ray diffraction (XRD) and reciprocal space mapping (RSM) data (Supporting Information, Fig. S2) confirm the coexistence of both antiferroelectric (O) and ferroelectric (R) phases in the LBFO thin film. The lattice parameters of R-LBFO and O-LBFO are a ≈ b ≈ 3.946 Å, c ≈ 3.961 Å, and a ≈ 3.955 Å, b ≈ 3.946 Å, c ≈ 3.915 Å, respectively, indicating full strain relaxation in the LBFO thin film.

Furthermore, the domain patterns of the two phases were characterized using piezoresponse force microscopy (PFM), as shown in Fig. 3. The topographic image of the LBFO thin film (Fig. 3a) reveals the coexistence of two distinct phases, exhibiting bright and dark contrasts, respectively. The corresponding out-of-plane and in-plane PFM images (Fig. 3b and 3c) show contrasts sensitive to ferroelectric polarization, revealing regions with high piezoelectric response (white and dark regions) and regions with zero piezoelectric response (dark-brown regions). The high piezoresponse regions correspond to the ferroelectric phase of the LBFO thin film, with the net polarization direction indicated by yellow arrows. In contrast, the regions with zero piezoresponse, marked by yellow dotted lines, correspond to the antiferroelectric phase.

This observation is further corroborated by cross-sectional transmission electron microscopy (TEM) images of the LBFO/SRO/DSO heterostructures (Fig. 3d and 3e), which show alternating light and dark bands on the same length scale as the stripes observed in the AFM images (Fig. 3f and 3g). From these TEM images, the spacing between the R-phase stripes is measured to be approximately 80 nm, while the O-phase stripes have a spacing of at least 120 nm. Corresponding line traces along the green lines in Fig. 3f and 3g show height changes of approximately 0.8 nm when transitioning from the R (bright) phase to the O (dark) phase. These results further confirm the coexistence of the antiferroelectric O phase and the ferroelectric R phase in the LBFO thin films.

To explore the phase transition behavior between the antiferroelectric (O) and ferroelectric (R) phases under an electric field, the phase transition behavior between the O and R phases under an electric field was examined using atomic force microscopy (AFM) and piezoresponse force microscopy (PFM). The AFM topography image in Fig. 4(a) and the out-of-plane PFM image in Fig. 4(e) reveal the coexistence of the R/O mixed-phase structure and the upward polarization direction in the initial state. In Fig. 4(a), the bright contrast corresponds to the R-phase LBFO, while the dark contrast stripes represent the O-phase LBFO. Upon applying a 16 V electric field within the dashed box, the stripe-like O-phase features vanish (Fig. 4(b)), indicating an electric-field-driven transition from the O phase to the R phase. Next, applying a negative voltage (-13 V) drives the R phase back to an R/O mixed-phase structure (Fig. 4(c)). Reapplying a positive voltage of 16 V once again enables reversible control of the R-O phase transition (Fig. 4(d)). The corresponding out-of-plane polarization switching is also demonstrated in Fig. 4(e)-(h). These results highlight the electric-field-driven, reversible antiferroelectric-to-ferroelectric phase transition in La-doped BFO thin films.

Although previous studies have reported that electric fields can irreversibly convert antiferroelectric phases into ferroelectric ones,[27-29] these transitions are often unstable and prone to returning to the initial antipolar state after a short time. This is generally considered a volatile saturation of antipolar dipoles along the direction of the applied electric field.[30,31] In contrast, the transition between the antiferroelectric and ferroelectric phases observed here is relatively stable. This stability can be attributed to the fact that the antiferroelectric phase is kinetically favored over the ferroelectric phase when a perpendicular electric field is applied.[32]

**CONCLUSIONS**

In summary, we demonstrate that gentle epitaxial strain induces a rhombohedral-to-orthorhombic phase transition in La-doped BFO thin films. Importantly, reversible O-R phase transitions driven by an electric field are also achieved. X-ray diffraction (XRD), reciprocal space mapping (RSM), and transmission electron microscopy (TEM) data confirm that the O-to-R phase transition is an antiferroelectric-to-ferroelectric phase transition. Our study paves the way for controlling the antiferroelectric-to-ferroelectric (O-R) phase transition to achieve magnetoelectric coupling in these thin films, enabling their potential use in related device applications.

## SUPPLEMENTARY MATERIAL

**See the supplementary material for details on structure and property of** the LBFO thin films.

## ACKNOWLEDGEMENTS

This work was supported by National Key Research and Development Programs of China (Grant No. 2022YFB3807603), the National Natural Science Foundation of China (Grant Nos. 92163210, 91963102, U1832104 and 52402146), Guangdong Basic and Applied Basic Research Foundation (2024B1515020076), Guangdong Provincial Key Laboratory of Optical Information Materials and Technology (No. 2017B030301007) and the Funding by Science and Technology Projects in Guangzhou (202201000008). Authors also acknowledge the financial support of Guangdong Science and Technology Project (Grant No. 2019A050510036), the Scientific Research Cultivation Fund for Young Faculty of South China Normal University (23KJ05) and Postdoctoral Fellowship Program of CPSF under Grant Number GZC20240519.

## AUTHOR DECLARATIONS

**Conflict of Interest**

The authors have no conflicts to disclose.

**Author Contributions**

## DATA AVAILABILITY

The data that support the findings of this study are available from the corresponding author upon reasonable request.

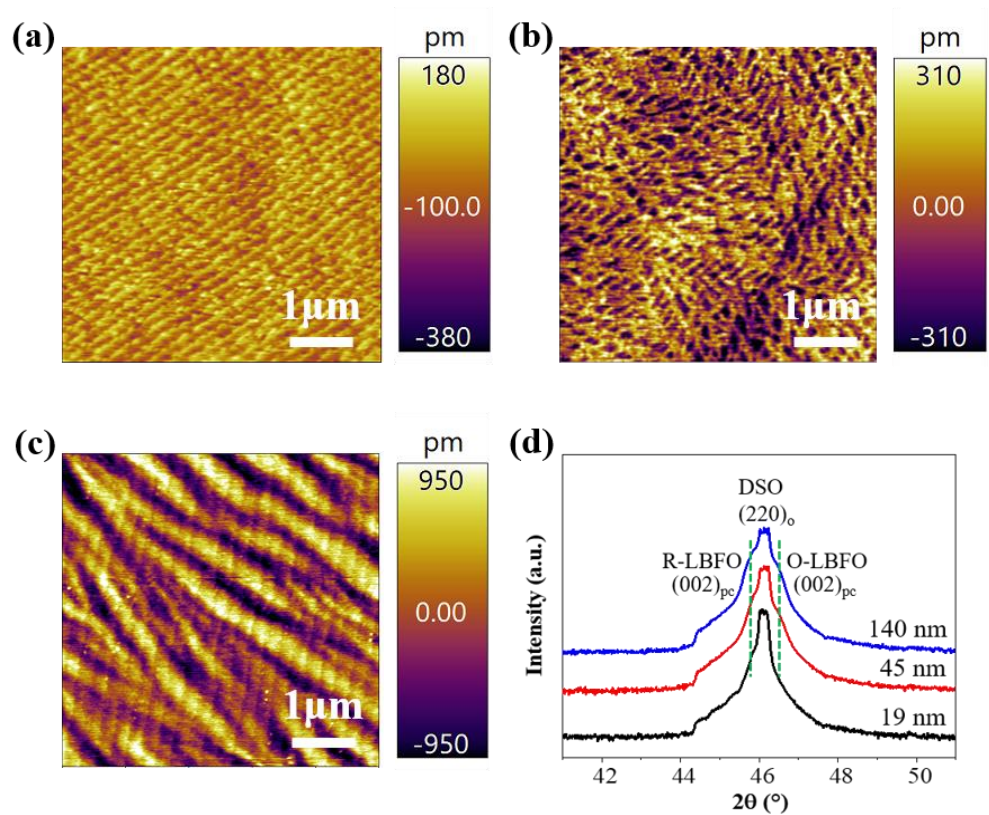

**FIG. 1.** Thickness-dependent evolution of surface morphologies as probed by AFM for (a) 19 nm, (b) 45 nm, and (c) 140 nm. (d) Typical XRD θ-2θ scans of LBFO thin films with different thicknesses grown on $(110)_O$-oriented DSO substrates.

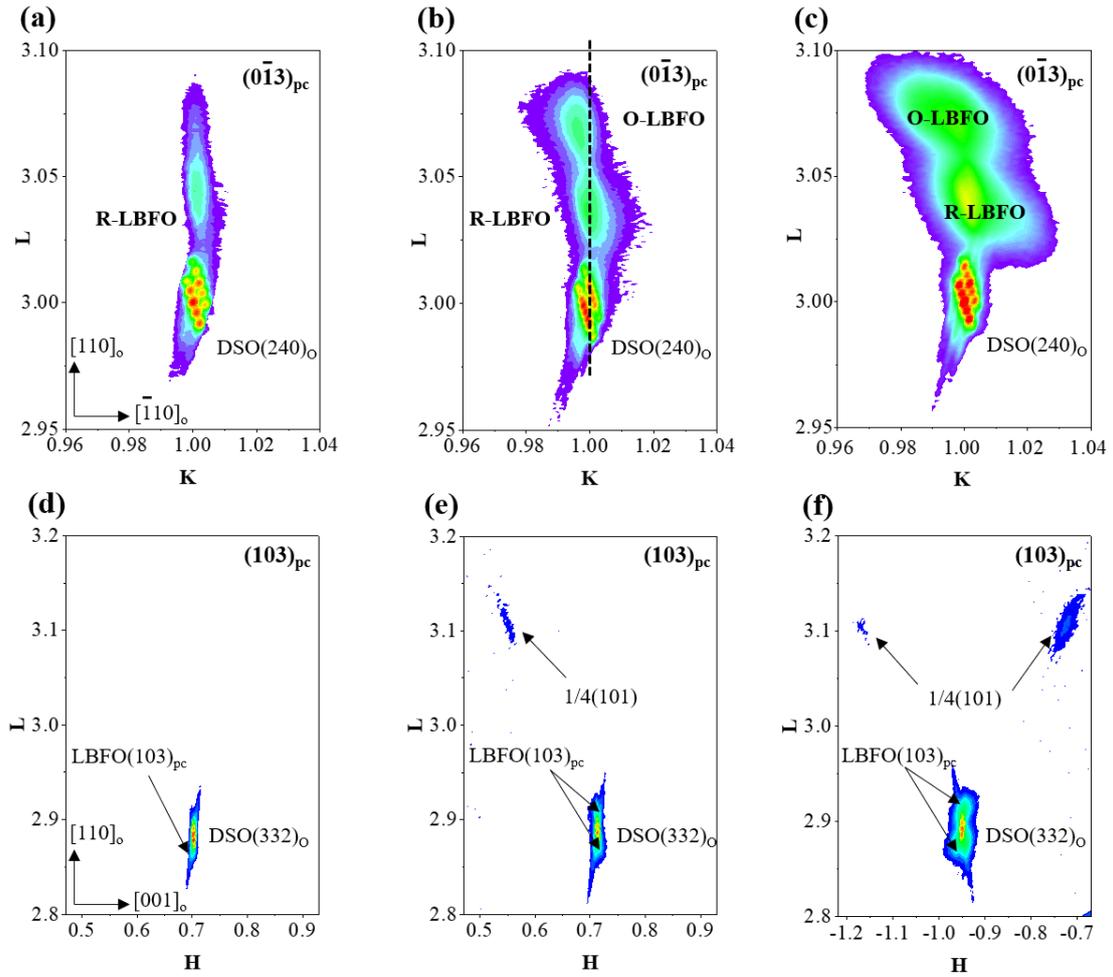

**FIG. 2.** RSMs of the (0-13) diffraction peaks of LBFO thin films with thicknesses (a) 19 nm, (b) 45 nm and (c) 140 nm. RSMs about the 1/4(101) diffraction conditions of the LBFO thin films with thicknesses (d) 19 nm, (e) 45 nm and (f) 140 nm are observed. The absence and emergency of a 1/4(101) diffraction spot confirms the ferroelectric-antiferroelectric phase transition.

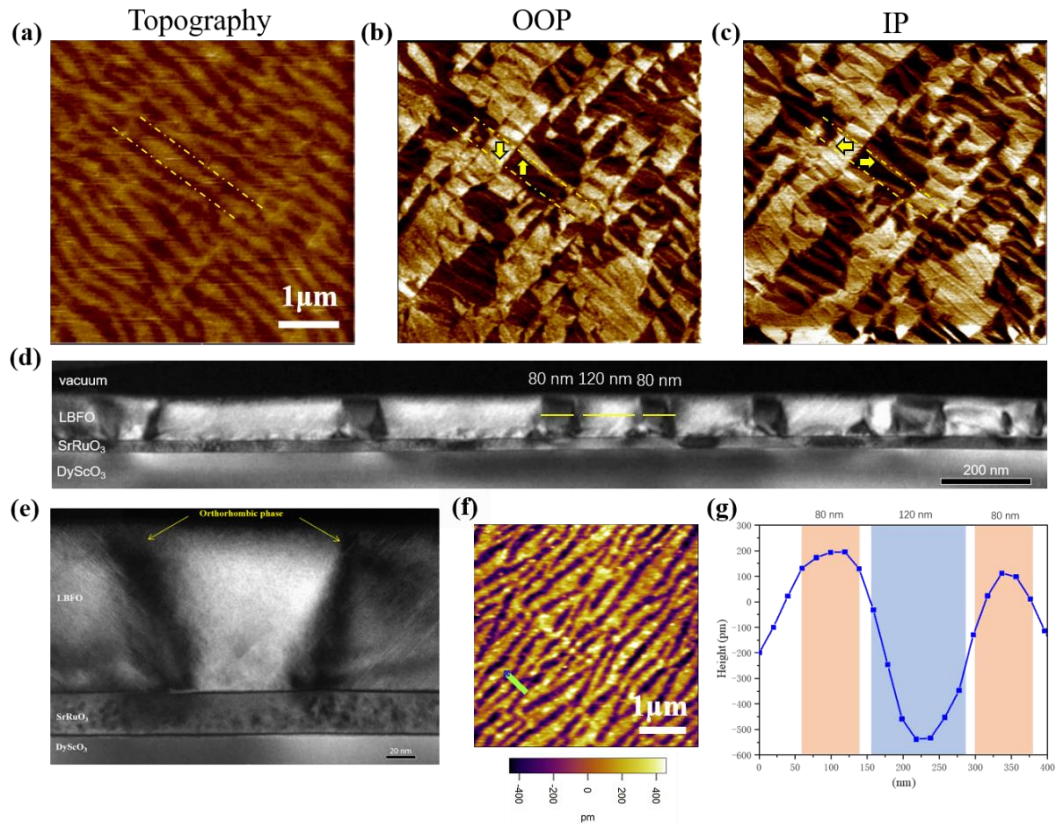

**FIG. 3.** (a) Topography, (b) out-of-plane, and (c) in-plane PFM images of LBFO thin film demonstrate the mixed O and R phase structures. The antiferroelectric O-phase dark stripes show a middle PFM contrast (dark brown regions and marked with yellow dotted lines) different from the ferroelectric R phase with the light and dark contrast (the net polarization direction marked with yellow arrows). (d) Dark-field large area cross-sectional TEM image of LBFO thin film. (e) Dark-field cross-sectional TEM image of a mixed O and R phase region in the LBFO thin film. (f) Topography across the multiferroic O-R phase boundaries (marked with green line). (g) Line profiles of the surface topography. The blue shadowed areas (corresponding to dark stripe-like region in Figure (f)) indicate the O-phase LBFO and yellow shadowed area indicate the R-phase LBFO, the width corresponding to the mixed O and R phase regions in Figure (d).

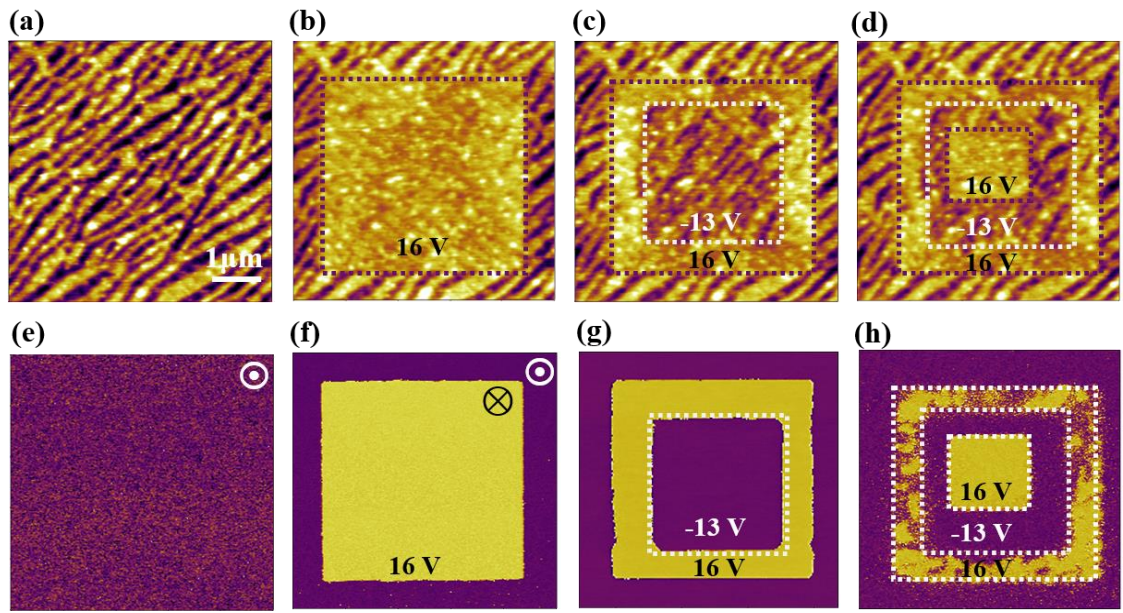

**FIG. 4.** Electrical control of phase transitions and polarization switching in O/R mixed phase LBFO thin film. (a)-(d) AFM images of 100 nm-thick LBFO with 20 nm-thick SRO on DSO substrate demonstrate the reversible O-R phase transition under electric field. (e)-(h) Out-of-plane PFM data corresponding to (a)-(d) show electrical switching of ferroelectric polarization.